# Machine Learning-Based Analysis of Ebola Virus' Impact on Gene Expression in Nonhuman Primates

Mostafa Rezapour, PhD[1*]; Muhammad Khalid Khan Niazi, PhD [1]; Hao Lu, PhD [1]; Aarthi Narayanan, PhD [2]; Metin Nafi Gurcan, PhD [1]

[1]Center for Artificial Intelligence Research, Wake Forest University School of Medicine, Winston-Salem, NC, USA
[2]Department of Biology, George Mason University, Fairfax, VA, USA

**\* Correspondence:**
Mostafa Rezapour, PhD
mrezapou@wakehealth.edu

## Abstract

This study introduces the Supervised Magnitude-Altitude Scoring (SMAS) methodology, a machine learning-based approach, for analyzing gene expression data obtained from nonhuman primates (NHPs) infected with Ebola virus (EBOV). We utilize a comprehensive dataset of NanoString gene expression profiles from Ebola-infected NHPs, deploying the SMAS system for nuanced host-pathogen interaction analysis. SMAS effectively combines gene selection based on statistical significance and expression changes, employing linear classifiers such as logistic regression to accurately differentiate between RT-qPCR positive and negative NHP samples. A key finding of our research is the identification of IFI6 and IFI27 as critical biomarkers, demonstrating exceptional predictive performance with 100% accuracy and Area Under the Curve (AUC) metrics in classifying various stages of Ebola infection. Alongside IFI6 and IFI27, genes, including MX1, OAS1, and ISG15, were significantly upregulated, highlighting their essential roles in the immune response to EBOV. Our results underscore the efficacy of the SMAS method in revealing complex genetic interactions and response mechanisms during EBOV infection. This research provides valuable insights into EBOV pathogenesis and aids in developing more precise diagnostic tools and therapeutic strategies to address EBOV infection in particular and viral infection in general.

## 1. Introduction

Ebola virus disease (EVD) is a severe and often fatal illness affecting humans and nonhuman primates (NHPs), with certain outbreaks, such as the 2013-2016 West Africa outbreak, resulting in high mortality rates [1, 2]. The scarcity of human clinical samples for research purposes has led to the reliance on NHP models, particularly cynomolgus and rhesus macaques, for studying EVD pathogenesis and testing potential treatments and vaccines. These models typically involve exposing animals to the Ebola virus (EBOV) through various methods, which produce symptoms and mortality patterns similar to those observed in humans, though differences in infection doses and disease progression are noted [1].

Furthering our understanding of the immune response to EVD, particularly through transcriptomic analyses, has been a significant outcome of studies using these NHP models [3, 4]. In this context, Speranza et al. [1] made a notable contribution by focusing on enhancing the accuracy of NHP models to more closely reflect human EVD. Their study involved the exposure of 12 cynomolgus macaques to the EBOV/Makona strain via intranasal routes, using a target dose of 100 plaque-forming units (PFU). The administration method varied, utilizing either a pipette or a mucosal atomization device. This led to a diverse onset of symptoms and disease progression among the animals, resulting in the identification of four distinct response groups with an overall fatality rate of 83% [1].

The animals were categorized into these groups based on the timing and nature of symptom appearance, onset of viremia, and time to death. Group 1, following a typical EVD course, showed quantifiable viremia from day 6 and had an average time to death of 10.47 days. Group 2, with a delayed onset, had detectable viremia between days 10 to 12 and an average time to death of 13.31 days. Group 3 experienced a late onset of disease, with detectable viremia emerging after day 20 and an average time to death of 21.42 days. Contrasting these, Group 4 did not develop detectable viremia during the experiment and survived until the 41-day study endpoint [1] (see **Figure 1**).

| Group 1 (Genome Equivalents/ml) | | | | |
|---|---|---|---|---|
| Day \ NHP | 0 | 3 | 6 | 10 |
| 1 | n.d. | n.d. | 8.60 | |
| 2 | n.d. | n.d. | 8.29 | 8.79 |
| 12 | n.d. | n.d. | 7.72 | 9.50 |

| Color | Description |
|---|---|
| n.d. | Not detectable (**Negative**) |
| | RT-qPCR Below Quantification Threshold |
| | **RT-qPCR–positive** |
| | **RT-qPCR–positive (GE>9: Strong Positive)** |

| Group 2 (Genome Equivalents/ml) | | | | | | | |
|---|---|---|---|---|---|---|---|
| Day \ NHP | 0 | 3 | 6 | 10 | 12 | 14 | 15 |
| 3 | n.d. | n.d. | | 9.68 | | | |
| 5 | n.d. | n.d. | n.d. | 9.21 | | 7.21 | 7.62 |
| 8 | n.d. | n.d. | | | X | | |
| 10 | n.d. | n.d. | | 10.04 | | | |

| Group 3 (Genome Equivalents/ml) | | | | | |
|---|---|---|---|---|---|
| Day \ NHP | 0 | 3 | 6 | 10 | 21 |
| 4 | n.d. | n.d. | n.d. | | 6.93 |
| 6 | n.d. | n.d. | | | |
| 9 | n.d. | n.d. | | | 9.81 |

| Group 4 (Genome Equivalents/ml) | | | | | | | | | |
|---|---|---|---|---|---|---|---|---|---|
| Day \ NHP | 0 | 3 | 6 | 10 | 14 | 21 | 28 | 35 | 41 |
| 7 | n.d. | n.d. | | | | | n.d. | n.d. | n.d. |
| 11 | n.d. | n.d. | n.d. | | | | | n.d. | n.d. |

*Figure 1.* The tables exhibit the timeline of responses for different nonhuman primates (NHPs) to EBOV exposure, arranged in rows, over a sequence of days post-exposure, which are organized in columns. Green cells indicate time points where reverse transcription quantitative polymerase chain reaction (RT-qPCR) did not detect any viral RNA (labeled as "n.d.", not detectable), suggesting a negative result. Yellow cells denote instances where RT-qPCR detected viral RNA, but the quantity was below the quantification threshold. Orange and purple cells highlight RT-qPCR-positive results, with purple specifically denoting a high viral load, as the Genome Equivalent (GE) is greater than 9, termed as "strong positive". The box marked with an "X" indicates that although the RT-qPCR results are positive, the corresponding NanoString information is unavailable.

In preparation for their experiment, Speranza et al. [1] conducted pre-exposure evaluations on blood chemistry, hematology, and soluble proteins, but these did not reveal any significant physiological differences among the groups, as confirmed by principal components analysis. Employing RNA sequencing and NanoString, which focuses on 769 NHP transcripts for transcriptomic analysis, they provided an overview of the host response to EBOV exposure. Their findings revealed a uniform and predictable response to lethal EVD, irrespective of the time to onset. Remarkably, they also discovered that the expression of specific genes could predict the development of disease before the appearance of clinical signs such as fever [1].

Despite the significant findings of Speranza et al. [1], certain aspects of EVD pathogenesis in NHPs remain underexplored, particularly in the context of genetic markers and their predictive power. While their study laid a solid foundation for understanding the disease's progression through transcriptomic analysis, it opened the door for more targeted research in identifying specific genetic indicators of EVD. This gap in knowledge presents an opportunity for employing advanced analytical techniques, such as machine learning, to delve deeper into the genetic landscape of EVD. Our study, therefore, focuses on expanding upon these initial findings, aiming to uncover finer genetic details that could be critical in diagnosing and treating EVD more effectively. By leveraging the rich dataset provided by Speranza et al. [1], we introduce a cutting-edge approach that promises to bring new insights into the complex interaction between the EBOV and its host at a molecular level.

Our research introduces a novel machine learning methodology aimed at prioritizing genes that are significantly associated with EVD, as determined by the Benjamin-Hochberg procedure [5]. This innovative approach allows us to not only identify gene fingerprints uniquely related to EVD but also to differentiate between EVD-positive and EVD-negative NHPs using a single gene at a time in a supervised manner. This dual focus on gene prioritization and individual gene-based differentiation represents a significant advancement in the field, offering a more nuanced understanding of the genetic underpinnings of EVD. Our methodology's precision in identifying and analyzing individual genes holds the potential to greatly enhance the specificity of EVD diagnostics and contribute to the development of targeted therapeutic strategies. Through this work, we aim to demonstrate the power of machine learning in uncovering new insights from existing genomic data, thereby opening new avenues for research in the field of infectious diseases.

## 2. Materials and Methods

### 2.1. Data:

This study utilizes a dataset of NanoString gene expression profiles from NHPs infected with the EBOV, as detailed in Speranza et al. [1]. The NanoString platform used recognizes 769 specific NHP transcripts, advantageous for its rapid processing capability and lower RNA quality requirements compared to RNA-seq, while retaining efficacy for EBOV research. Normalization of the NanoString data was conducted in line with standard procedures, where background adjustments were made based on negative controls, and lane variations were accounted for using internal positive controls [1]. The most stable reference genes for normalization were identified using the NormFinder R package [6], ensuring precise calibration for RNA input variations. Further methodological specifics are available in the referenced work by Speranza et al. [1], which should be consulted for an in-depth understanding of the protocols and procedures applied.

### 2.2. Supervised Magnitude-Altitude Scoring (SMAS): A Machine Learning Approach for Gene Expression Profiling

This section describes the Supervised Magnitude-Altitude Scoring (SMAS) method, specifically tailored to distinguish RT-qPCR positive and negative NHP samples in EBOV research. SMAS comprises a structured three-stage process:

1. In the initial stage of gene selection, our methodology rigorously identifies genes that demonstrate statistical significance between RT-qPCR positive and negative groups. To address the challenges of multiple hypothesis testing inherent in gene expression studies, we strictly apply the Benjamini-Hochberg (BH) correction [7]. This correction is essential for controlling the false discovery rate, a critical factor in genomic data analysis where numerous tests are conducted simultaneously. We employ two-sample independent t-tests [8] with the BH correction, ensuring that identified genes are significantly different beyond random chance occurrences. Moreover, our criteria for selecting genes include a log fold change (logFC) greater than 1. This means we only consider genes that not only pass the BH significance threshold but also exhibit substantial expression differences.

2. In the second stage of our analysis, the Magnitude-Altitude Score (MAS) is calculated for each gene selected in the initial stage. The Magnitude-Altitude Scoring (MAS) formula is:

$$MAS_l = |(\log 2FC_l)|^M |(\log_{10}(p_l^{BH}))|^A,$$

for $l = 1, 2, \ldots, s$, where $s$ is the number of rejected null hypotheses by BH adjusted method ($p_l^{BH} < \alpha$, where $\alpha = 0.05$). The hyperparameters M and A are used to achieve a balance between the adjusted p-value and the log fold change, facilitating a comprehensive analysis of gene expression changes. In traditional gene ranking methods employed by differential expression analysis tools like EdgeR [9] (or DESeq2 [10]), the focus is predominantly on p-values for ranking genes. This approach is exemplified in EdgeR's typical workflow, where genes are ranked using the "topTags" function. This function sorts genes primarily based on their p-values, which are calculated to determine the statistical significance of differential expression between conditions. While this method effectively identifies statistically significant genes, it may not always highlight genes with the most biologically significant changes in expression.

In contrast, our MAS system, with both M and A set to 1, goes beyond traditional p-value ranking. By incorporating the log fold change (logFC) into the ranking process alongside the adjusted p-values from the Benjamini-Hochberg correction, MAS makes it possible that the genes selected for further analysis show not only statistical significance but also potential biologically relevant expression changes. This balanced approach of considering both statistical and biological significance allows for a more comprehensive understanding of gene expression changes in the context of EBOV infection.

Genes are ranked according to their MAS, and the top-$d$ genes are earmarked for predictive modeling. The number d of genes selected is capped at one-tenth of the total number of samples, in line with a widely accepted machine learning guideline that suggests having at least ten samples for each variable. This guideline helps prevent overfitting and possibly increases the robustness of the model [11]. By adhering to this principle, our approach effectively balances the inclusion of a sufficient number of genes to capture the EBOV infection's complexity while maintaining robustness.

3. In the final stage of our methodology, the SMAS approach demonstrates its efficacy through the implementation of linear classifiers for EVD status prediction. The top-d genes, meticulously identified through MAS, serve as predictors in linear classifiers, specifically employing logistic regression or support vector classifiers with a linear kernel [12]. This choice of linear classifiers is strategically made considering the often-limited sample size encountered in multi-omic datasets. In such scenarios, more complex models are prone to overfitting, potentially skewing predictions [11]. By opting for linear classifiers, we effectively mitigate this overfitting risk, striking a balance between model simplicity and the ability to accurately predict EVD status in unseen NHP samples.

To assess the robustness of our predictive model, we implement K-fold stratified cross-validation [13]. This validation technique is critical in our study as it ensures a balanced representation of both positive and negative samples in the training and testing datasets. Such a balanced approach is pivotal in enhancing the reliability and generalizability of our model. By incorporating a cross-validation strategy, we are not only able to gauge the model's performance on unseen data but also affirm its stability and predictive power across different subsets of the data.

Integrating the top-$d$ genes from the SMAS method into linear classifiers showcases the utility of our approach in practical applications. By selecting genes that are both statistically significant and biologically relevant and deploying them in a well-validated predictive model, the SMAS method presents a potent tool in the field of infectious disease research, particularly for EVD analysis. This stage, therefore, underscores the effectiveness of the SMAS method in translating complex genomic data into actionable insights, paving the way for advancements in diagnostic and therapeutic strategies.

Through the SMAS method's comprehensive stages, we chose to set both M and A to 1 in the MAS formula, as opposed to the traditional ranking method where $M = 0$ and $A = 1$, which reduces the analysis to the adjusted p-values from the Benjamini-Hochberg correction. This decision was based on the following rationale:

1. Enhanced discrimination between groups: The MAS scoring with $M = A = 1$ takes into account both the statistical significance and the magnitude of change in gene expression. This dual consideration allows for a more nuanced differentiation between the RT-qPCR positive and negative groups. In contrast, the traditional ranking method primarily focuses on statistical significance, potentially overlooking genes with substantial biological changes.

2. Biological relevance of log fold change (logFC): By setting a threshold for logFC (e.g., $logFC > 1$), we increase the possibility that genes with biologically significant expression changes are considered. This threshold is crucial as it highlights genes that are not just statistically significant but also have substantial changes in expression levels, which is more likely to be biologically relevant in the context of EBOV response in NHPs.

3. Improved predictive performance in supervised learning: In supervised machine learning, the balance between statistical significance and expression magnitude is vital for model accuracy. The MAS scoring with $M = A = 1$, combined with the logFC threshold, tends to select genes that provide a clearer separation between positive and negative samples. This improved separation enhances the predictive performance of the model, particularly when using linear classifiers like logistic regression or support vector classifiers with a linear kernel.

In summary, adopting MAS scoring not only aligns with the principles of robust machine learning but also ensures that the genes selected for the predictive model are the most informative for distinguishing EVD status in NHP samples.

## 2.3. Differential Expression Analysis and Classification Using Supervised Magnitude-Altitude Scoring (SMAS)

In our exploration of the EBOV's interactions with its host, we have developed a methodical approach to analyze the dynamics of viral infection. This approach is segmented into two primary objectives:

**Objective 1:** a comprehensive differential expression analysis and

**Objective 2:** the development of a bi-class classification model.

**Figure 2** illustrates our structured approach to studying EBOV-host interactions, breaking down into two primary objectives. The first main objective consists of four subobjectives, each dedicated to a detailed differential expression analysis. In these subobjectives, we focus on genes that are significant according to the Benjamini-Hochberg method, exhibit a log fold change (logFC) greater than 1, and are further prioritized using MAS method to delineate the gene expression changes at different stages of EBOV infection in NHPs. The collective findings from these subobjectives aim to provide a comprehensive understanding of the host's genetic response. The second main objective leverages the key genes identified in Objective 1 to develop a bi-class classification model, utilizing these genes as predictors to classify NHP samples as either positive or negative for EBOV infection.

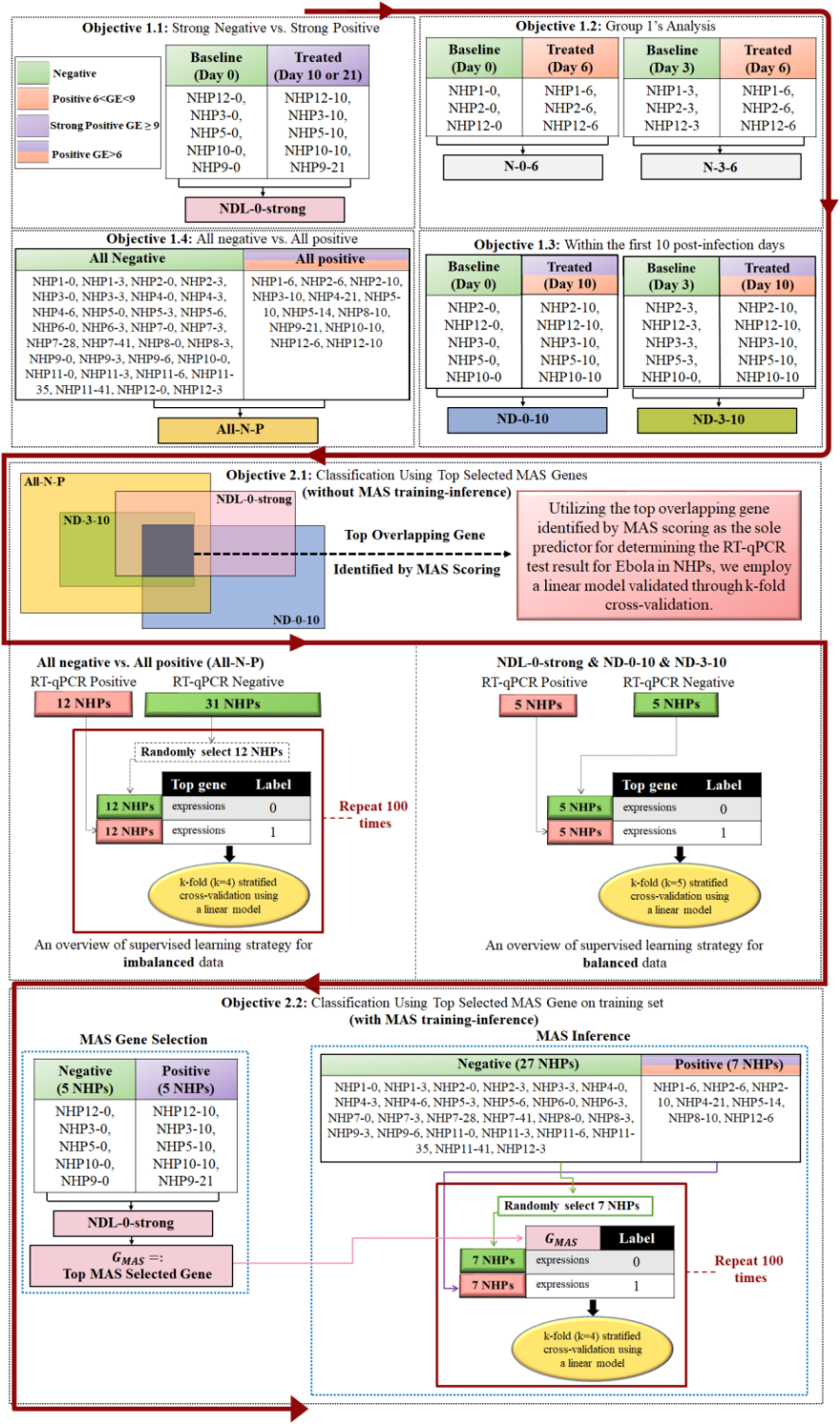

***Figure 2.*** *This figure encapsulates the structured approach of Objective 1, which is divided into four subobjectives, each conducting differential expression analyses on NHP samples at various stages of EBOV infection. Genes are selected based on Benjamini-Hochberg significance with logFC > 1 and prioritized using the MAS method. The naming convention "NHP**m-n**"*

refers to NanoString data related to nonhuman primate number "*m*" on day "*n*" post-infection, as indicated in Figure 1. The integrated results of these analyses are then applied in Objective 2 to build a bi-class classification model for predicting EBOV infection status. In Objective 1.4, since NanoString data for NHP8-12 was not available, but NHP8-10 data was available with three RT-qPCR replicates that were positive, we used NHP8-10 as the positive sample.

**Objective 1. Comprehensive Differential Expression Analysis Using MAS Scoring:**

Objective 1 systematically dissects the impact of the EBOV on gene expression in NHPs, utilizing the MAS system. This objective is divided into four key sub-objectives, each targeting a specific aspect of the host-pathogen interaction, thus providing a multi-layered understanding of the EBOV's genetic impact. Objective 1.1 focuses on extreme cases of infection, comparing the most intense positive and negative responses. This specific analysis is critical for identifying genetic markers that undergo significant alterations in severe infection scenarios, revealing key genes that are most responsive or vulnerable during heightened viral replication. Objective 1.2 then shifts focus to the early stages of infection, examining gene expression dynamics from initial exposure to the peak of viral response in Group 1. This objective is instrumental in understanding the progression of early-response genes, providing crucial insights for early detection and intervention.

Objectives 1.3 and 1.4 further expand the analysis, each adding a vital layer of understanding. Objective 1.3 explores gene expression in typical and delayed response scenarios, highlighting the genetic basis for variability in host response timing. This objective is particularly important for understanding genes involved in delayed viral response or clearance, offering valuable information on how different hosts react to the virus over time. Finally, Objective 1.4 provides a broad overview, comparing all negative and positive infection statuses. This comprehensive comparison captures the overall genetic signature of EBOV infection, identifying consistent genetic markers across various infection stages. Together, these objectives offer a multi-layered and detailed understanding of the host's genetic response to the EBOV, ensuring a thorough exploration of pathogenesis and host response at the genetic level. This structured approach is crucial for revealing the complex genetic landscape shaped by EBOV infection, contributing significantly to developing more precise diagnostic tools and therapeutic strategies.

**Objective 1.1. Strong Negative vs. Strong Positive Analysis (NDL-0-strong):**

This objective engages in a detailed analysis to explore the multifaceted impact of the EBOV on gene expression in NHPs, employing the MAS system. The focus is to dissect the genetic response in NHPs across varying degrees of EBOV infection intensity, thereby illuminating the broad spectrum of host-pathogen interactions.

Within this comprehensive framework, a specialized subset of analysis, termed as "Strong Negative vs. Strong Positive Analysis (NDL-0-strong)," is conducted. This segment specifically aims to distinguish the gene expression contrasts between the most extreme cases of EBOV infection. By comparing the gene expression profiles of strong positive RT-qPCR samples (indicated by a high Genome Equivalent, GE>9, and marked as purple in **Figure 1**) against those of strong negative samples (from day 0), this analysis provides a deep dive into the genetic shifts occurring under intense viral influence.

The procedure involves carefully selecting samples from Groups 1, 2, and 3 (see **Figure 1**) that exhibit strong positive RT-qPCR results. These samples are then methodically contrasted with the baseline samples taken on day 0, focusing particularly on NHPs that later exhibit strong positive outcomes. This approach allows for a targeted analysis of significant gene expression changes associated with high GE values. For clarity and ease of reference, the set of all genes selected using the MAS scoring with a log fold change (logFC) greater than 1 in this phase is collectively referred to as "NDL-0-strong."

The rationale behind this intense focus lies in the value of contrasting extreme cases: strong positive (high GE) versus strong negative (day 0) samples. Such an approach is instrumental in unearthing insights into the genetic markers most drastically altered in the wake of severe EBOV infection. This comparison not only aids in understanding the genetic extremes induced by the virus but also plays a crucial role in highlighting the key genes that are most responsive or vulnerable during heightened viral replication phases. Through this nuanced analysis, the study endeavors to unravel the complex genetic landscape shaped by EVD, thereby contributing significantly to the broader understanding of the virus's genetic impact on its hosts (see **Figure 2**).

**Objective 1.2. Differential Expression in Group 1 (N-0-6 & N-3-6):**

Objective 1.2 delves into the specific gene expression dynamics of Group 1 (see **Figure 1**), which is characterized by a typical EBOV infection progression. This analysis focuses on understanding the transition from the initial exposure to the peak viral response, primarily observed on day 6 post-exposure.

The method involves applying the MAS within Group 1 to distinguish significant gene expression changes between day 6 (when RT-qPCR is positive) and days 0 and 3 (when RT-qPCR is negative). The genes with a log fold change (logFC) greater than 1, when comparing day 6 against day 0, are categorized as "N-0-6", while those compared against day 3 are labeled "N-3-6" (see **Figure 2**).

This approach is critical for identifying genes that respond early or undergo significant alterations during the acute phase of the infection. By examining these key time points, we can pinpoint the genetic markers that are pivotal in the initial stages of viral replication and host response.

**Objective 1.3. Analysis Within Normal and Delayed Groups (ND-0-10 & ND-3-10):**

Objective 1.3 aims to unravel the gene expression patterns in scenarios of typical and delayed EBOV response (Groups 1 and 2 in **Figure 1**), particularly noticeable by day 10 post-infection. This objective targets an understanding of the genetic underpinnings that might influence the timing of viral detection and clearance.

The procedure employs MAS scoring to analyze gene expression in normal and delayed response groups with positive RT-qPCR results on day 10. The set of all genes selected with a log fold change greater than 1 when comparing day 10 against day 0 is referred to as "ND-0-10", and the gene set from the comparison of day 10 against day 3 is termed "ND-3-10" (see **Figure 2**).

This analysis is vital in exploring genes that could be responsible for a delayed response to the EBOV, thereby providing insights into the genetic basis for variability in host response timing. It helps to understand how different gene expression patterns contribute to the delayed manifestation of the virus.

**Objective 1.4. Comparison of All Negative vs. All Positive Samples (All-N-P):**

Objective 1.4 encompasses a broader perspective, aiming to distinguish the overarching gene expression signatures that separate overall negative from positive EBOV infection statuses. This comprehensive comparison seeks to capture the complete genetic landscape associated with EBOV presence or absence.

Using MAS scoring, this analysis compares the collective gene expression of all negative samples (indicated as green in **Figure 1**) against all positive samples (denoted as orange and purple). This broad comparison is instrumental in identifying markers consistently expressed across various stages of infection, categorized collectively as "All-N-P."

The rationale behind this objective is to provide a holistic view of the gene expression changes associated with EBOV infection. It is designed to capture the overall genetic signature indicative of infection, thereby differentiating between infected and uninfected states in a comprehensive manner (see **Figure 2**).

Objectives 1.1 to 1.4 provide a layered and detailed understanding of the host's genetic response to the EBOV. These objectives collectively aim to identify gene fingerprints that can effectively differentiate between NHPs testing positive and those testing negative for RT-qPCR across varying infection scenarios and response patterns. This structured approach ensures a multifaceted analysis, contributing significantly to the broader understanding of EBOV pathogenesis and host response at the genetic level.

**Objective 2. Application of MAS Insights for EBOV Classification**

The primary goal of our paper is to identify gene signatures associated with EBOV infection by employing differential expression analysis through the MAS method. This unsupervised machine learning approach, which forms the crux of Objective 1, operates without traditional training methods and is instrumental in uncovering significant gene expressions linked to the EBOV. Objective 2 builds on this foundation, extending our analysis from gene identification to practical application in EBOV research. This objective bifurcates into two distinct yet interconnected components.

In Objective 2.1, our focus shifts to utilizing linear classifiers to demonstrate the efficacy of genes selected through the MAS method. Here, we illustrate how these MAS-selected genes can effectively differentiate between RT-qPCR negative and positive samples of EBOV in nonhuman primates. This stage is pivotal in showcasing how the insights gained from unsupervised machine learning can be transitioned into practical, diagnostic applications.

In Objective 2.2, we demonstrate the MAS method's capacity for generalization by selecting a gene from one independent dataset and then applying it for supervised classification on a different dataset. This approach not only tests the selected gene's predictive power but also showcases the adaptability and robustness of the MAS method across diverse contexts. By employing the MAS-identified gene in a separate dataset to accurately classify EBOV infection status, we provide compelling evidence of the MAS's generalizability. The ability of the MAS method to extend its application from gene selection in one scenario to effective prediction in another exemplifies its utility and versatility in infectious disease research, especially in studies of complex viral infections like Ebola.

**Objective 2.1. Classification Using Top Selected MAS Genes:**

Objective 2.1 capitalizes on the insights gained from the differential expression analysis in Objective 1 to develop a binary classification model. This model distinguishes between positive and negative EBOV infection statuses in NHPs based on their gene expression profiles. Utilizing logistic regression, the model integrates the most significant genes identified across all sub-objectives of Objective 1 as predictive variables. To validate and ensure the reliability of this model, we employ a k-fold stratified cross-validation approach, which rigorously assesses its performance and generalizability.

The core rationale of Objective 2.1 is to transform the complex gene expression data into a practical diagnostic tool. Leveraging the key genes identified through the MAS method, the model aims to provide a robust and precise means of classifying EBOV infection status. This application of genomic insights into a functional diagnostic model not only underscores the utility of our analytical approach but also has the potential to inform targeted therapeutic interventions.

A significant aspect of Objective 2.1, as illustrated in **Figure 2**, involves addressing the challenge of imbalanced datasets, a common issue in biomedical research. For instance, in the All-N-P group analysis, we encounter an imbalance with 12 positive samples and 31 negative samples. To tackle this, we implement a random sampling strategy: selecting 12 negative samples to pair with the 12 positive samples, thereby creating a balanced dataset of 24 samples. This process is repeated 100 times, each with a different subset of negative samples, ensuring a comprehensive evaluation. The performance of the model across these iterations is assessed through metrics such as the average Area Under the Curve (AUC), accuracy, precision, recall, and F1-score. This approach not only enhances the model's robustness but also provides a more accurate reflection of its predictive capability.

In summary, Objective 2.1, in conjunction with Objective 1, forms a holistic analysis of the EBOV's genetic interactions with its host. While Objective 1 establishes the foundation by pinpointing critical genes and their expression patterns in response to EBOV infection, Objective 2.1 translates these findings into a pragmatic classification tool. This synergistic blend of in-depth genomic analysis and practical application is pivotal in advancing the field of infectious disease research, particularly in the context of EBOV diagnostics and therapy.

**Objective 2.2. Implementing MAS on an Independent Set for Subsequent Supervised Classification:**

Initially, MAS is applied to a balanced dataset, characterized by strong positive contrasts against strong negatives, acting as a basis for gene selection. The top MAS-selected gene, named $G_{MAS}$, is then identified. $G_{MAS}$ is subsequently used as the sole predictor in a linear classifier to differentiate RT-qPCR positive and negative NHP samples in a separate dataset, which remains unexposed during the MAS top gene selection process.

As shown in **Figure 2**, the necessity for strategic MAS application arises from the limited sample size. MAS is employed on the ND-0-strong subset to isolate the most significant gene (top-MAS selected gene) for further examination. This identified gene, $G_{MAS}$, becomes central to subsequent classification tasks across different sample sets. These tasks are thoroughly validated through balanced stratified cross-validation techniques as shown in **Figure 2**.

In essence, Objective 2 is where the theoretical and analytical advancements of our study converge into practical applications. It showcases the transition from identifying gene signatures associated with EBOV infection using an unsupervised machine learning approach, to applying these insights in a supervised setting, thus advancing the field of infectious disease research and offering new avenues for EBOV diagnostics and treatment strategies.

# 3. Results

In this section, we present the results of the methodology outlined in Section 2.

**Objective 1. Comprehensive Differential Expression Analysis Using MAS Scoring:**

For objectives with at least 10 NHPs, we conducted supervised analysis to predict whether an NHP is RT-qPCR positive or negative.

**Objective 1.1. Strong Negative vs. Strong Positive Analysis (NDL-0-strong):**

**Figure 3** displays the results of analysis for Objective 1.1, where we compare the NHPs with strong positive RT-qPCR results against their status on day 0, when RT-qPCR results were strongly negative. **Figure 3(a)** depicts the volcano plot and highlights the top MAS-selected genes with a log fold change (logFC) greater than 1. **Figure 3(b)** presents a three-dimensional visualization of all strongly positive and negative NHPs, utilizing only the top three selected MAS genes: IFI27, IFI6, and HP. Given that we have more than 10

NHPs for this objective, we implemented the Supervised Magnitude-Altitude Scoring (SMAS) approach (Objective 2) using logistic regression. This was carried out with a 5-fold stratified cross-validation, specifically focusing on the top selected gene, IFI27. The Receiver Operating Characteristic (ROC) curves for each of the five folds, along with their mean, are depicted in **Figure 3(c)**. This figure also includes the AUC and accuracy metrics for each fold, providing a comprehensive evaluation of the model's predictive performance. It is important to note that we opted for a single gene and a linear model to ensure the model's strong generalizability, as the MAS-selected gene IFI27 is exceptionally suitable for this purpose.

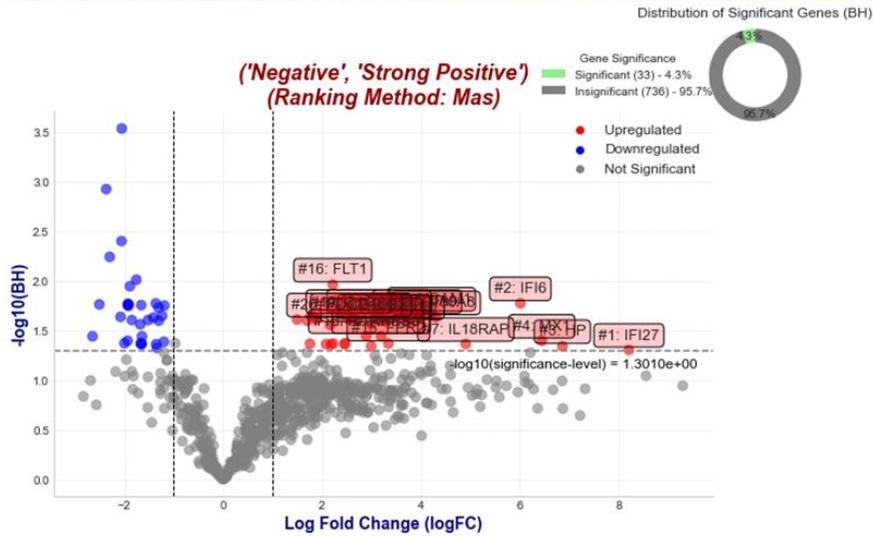

Figure 3(a)

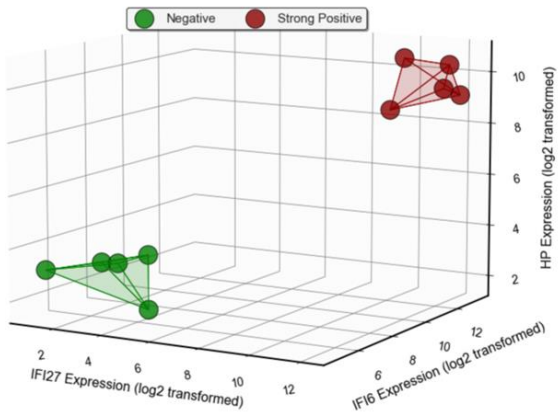

Figure 3(b)

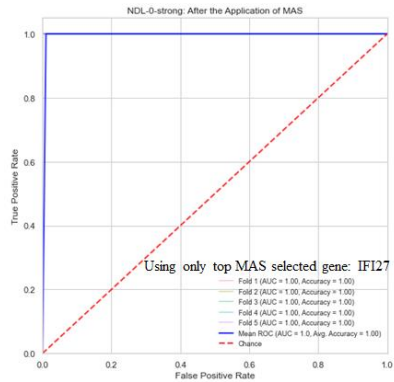

Figure 3(c)

*Figure 3.* Analysis and Visualization of Gene Expression and Predictive Modeling in Objective 1.1: Figure 3(a) features a volcano plot that contrasts gene expressions of NHPs with strong positive RT-qPCR results against their status on day 0, when RT-qPCR results were strongly negative, highlighting top MAS-selected genes with a logFC greater than 1. In Figure 3(b), a three-dimensional visualization is provided, showing all strongly positive and negative NHPs using the top three selected MAS genes: IFI27, IFI6, and HP. Furthermore, Figure 3(c) showcases the Receiver Operating Characteristic (ROC) curves for each of the five folds of the 5-fold stratified cross-validation, along with their mean, using logistic regression on the top selected gene, IFI27. This panel also includes AUC and accuracy metrics for each fold, offering a detailed assessment of the predictive model's

*performance. The focus on a single gene and linear model emphasizes the model's robust generalizability, demonstrating the effectiveness of IFI27 as a predictive marker.*

**Objective 1.2. Differential Expression in Group 1 (N-0-6 & N-3-6):**

**Figure 4** illustrates the results of Objective 1.2. **Figures 4(a)** and **4(b)** display the volcano plots and Benjamini-Hochberg (BH) significant genes when comparing NHPs from day 6 to day 0 and day 3, respectively. IFI27, IL6, and THBD are the only BH-significant genes with a log fold change (logFC) greater than 1 when comparing NHPs on day 6 versus day 0. However, when comparing day 6 to day 3, only IFI27 remains significant. Notably, for the purpose of distinguishing NHPs on day 3 from day 0, we also performed MAS for Day 3 versus Day 0, but no gene emerged as BH significant (see **Figure 4(c)**). Nevertheless, when relaxing the Benjamini-Hochberg criteria and considering only raw p-values, FLT1 emerged as the top significant gene with a logFC greater than 1 based on MAS (see **Figure 4(d)**). **Figure 4(e)** illustrates the expression of IFI27 (selected from Objectives 1.1 and 1.2) and IFI6 (selected from Objective 1.1), along with FLT1, which distinguishes NHPs on day 0 from day 3. Since we have a total of 9 NHPs across all three post-infection days, we did not perform a supervised learning analysis.

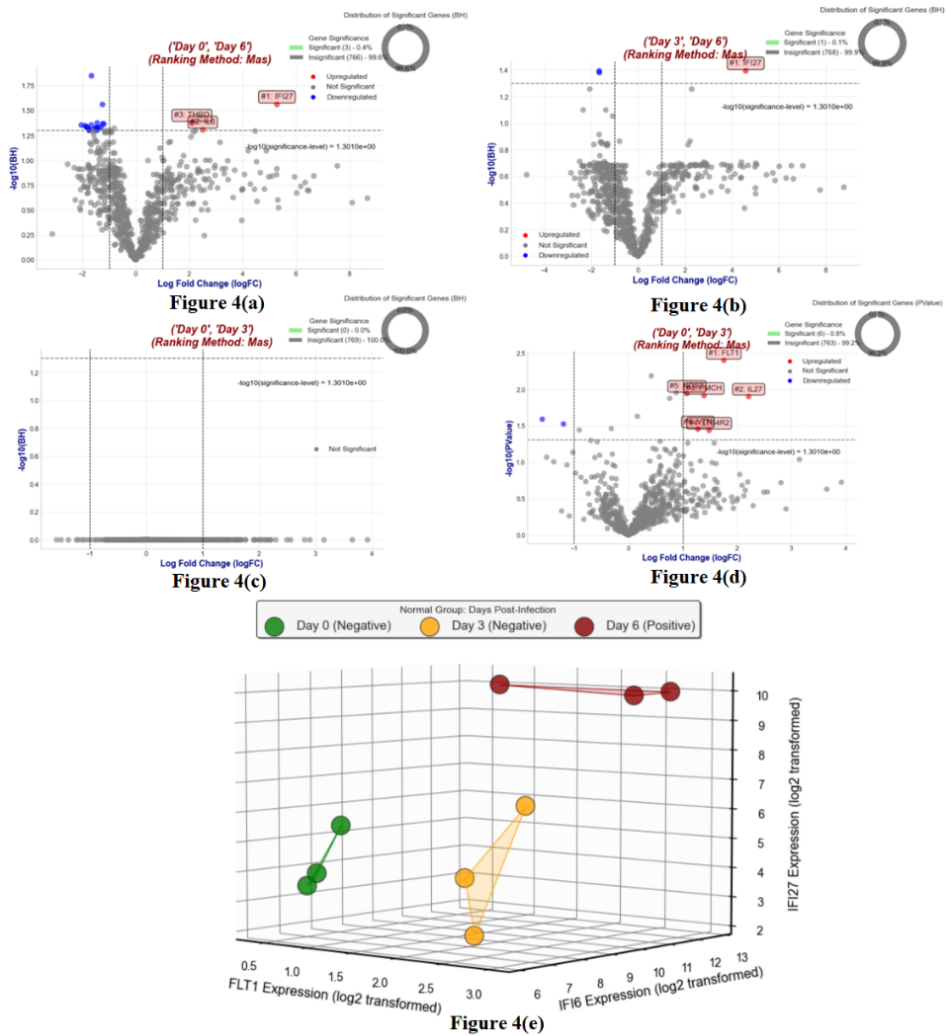

*Figure 4. Comprehensive Overview of Gene Expression Analyses in Objective 1.2: Figure 4(a) displays a volcano plot comparing gene expressions of NHPs on day 6 versus day 0, highlighting significant changes. Figure 4(b) shows a similar plot for day 6 versus day 3, pinpointing early infection response genes. Figure 4(c) illustrates the lack of BH significant genes when*

*comparing day 3 to day 0. Figure 4(d) reveals FLT1 as a top significant gene with relaxed BH criteria. Finally, Figure 4(e) compares the expression patterns of IFI27 (from Objectives 1.1 and 1.2), IFI6 (from Objective 1.1), and FLT1, emphasizing their roles across different infection stages.*

**Objective 1.3. Analysis Within Normal and Delayed Groups (ND-0-10 & ND-3-10):**

**Figure 5** presents the results corresponding to Objective 1.3. **Figure 5(a)** displays a Venn diagram that illustrates the BH significant genes with a logFC greater than 1 when comparing NHPs on day 10 versus day 0 and day 10 versus day 3, highlighting both unique and common genes prioritized by MAS. ISG15, IFI6, and IFI44 emerge as the top three selected genes significantly expressed in both comparisons. Notably, the absence of unique significant genes in the day 10 versus day 3 comparison suggests an early expression of these genes. **Figures 5(b)** and **5(c)** depict the unique and common BH significant genes across Objectives 1.1, 1.2, and 1.3. From **Figure 5(b)**, the presence of IFI27 as the sole BH significant gene with a logFC greater than 1 in the ND-0-10, N-0-6, and NDL-0-strong comparisons is noteworthy. **Figure 5(c)** indicates that IFI6, HP, and IFI27 are the top three BH-significant genes with a logFC greater than 1, common among ND-0-10, ND-3-10, and NDL-0-strong. Utilizing IFI6, IFI27, or ISG15 (only one gene) as the sole predictor in SMAS, **Figure 5(d)** displays the ROC AUC for classifying NHPs with negative RT-qPCR on day 0 versus positive ones on day 10. **Figure 5(e)** shows a similar analysis for classifying samples as positive on day 10 and negative on day 3.

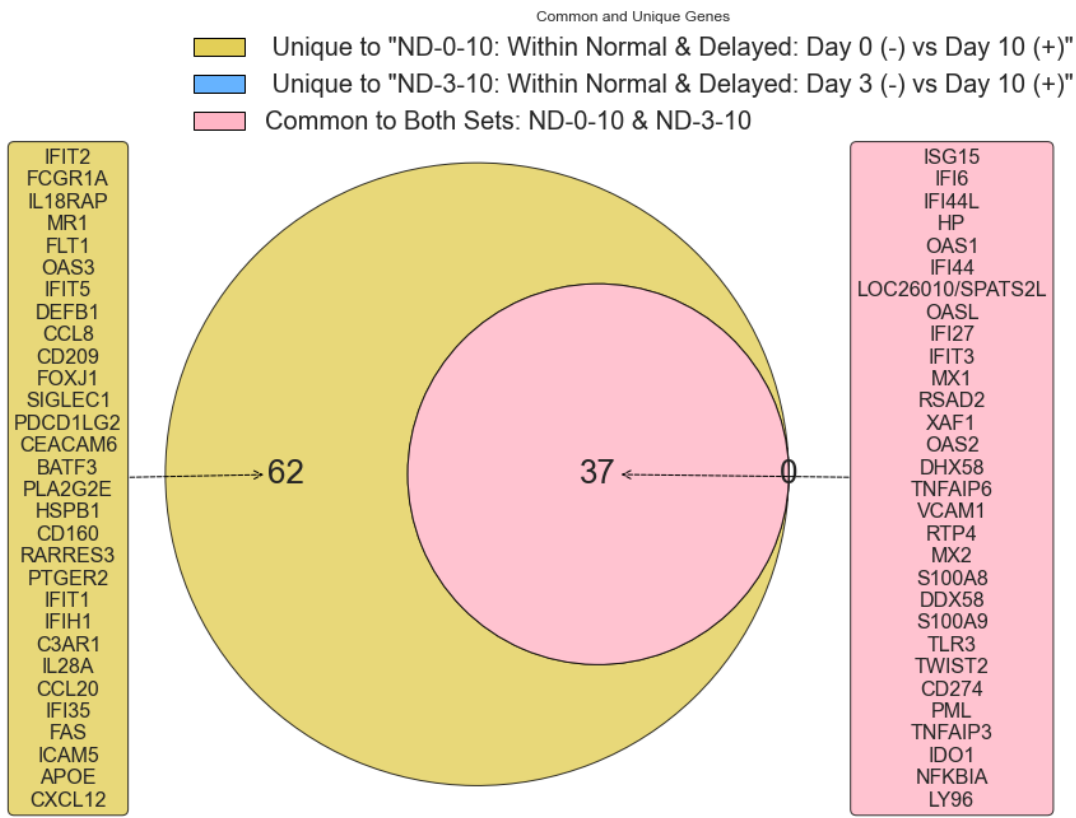

***Figure 5(a).*** *Venn Diagram of BH Significant Genes - This diagram compares NHPs on day 10 versus days 0 and 3, showcasing BH significant genes with logFC > 1. It highlights the top genes (ISG15, IFI6, IFI44) expressed significantly in both contrasts and the lack of unique significant genes for the day 10 vs day 3 comparison.*

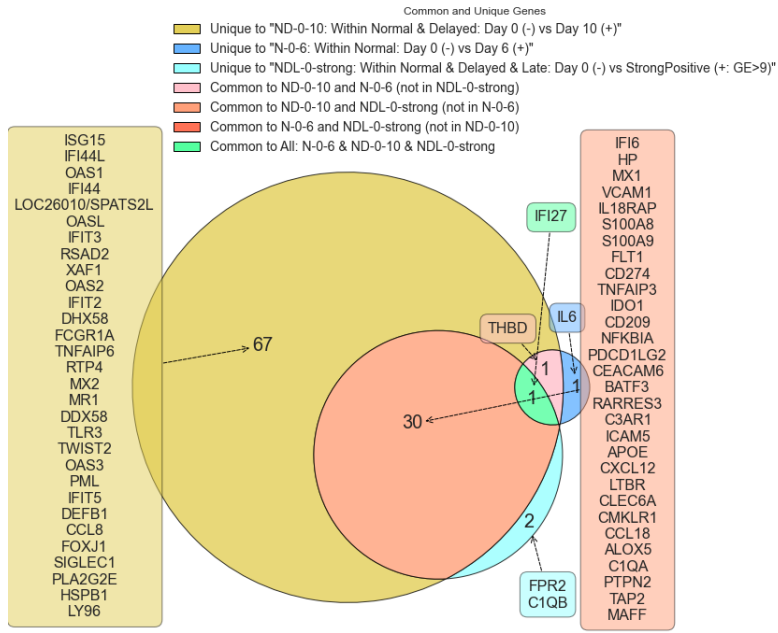

***Figure 5(b).*** *Unique BH Significant Genes Across Objectives - This figure reveals IFI27 as the only BH significant gene with logFC > 1 when comparing ND-0-10, N-0-6, and NDL-0-strong, underscoring its unique presence across different comparisons.*

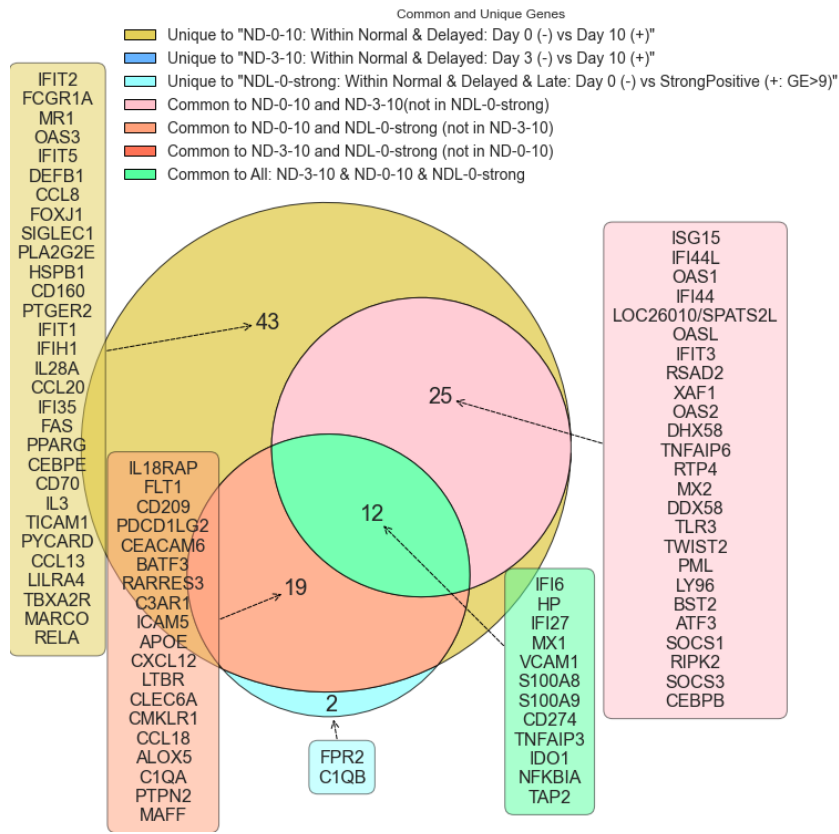

***Figure 5(c).*** *Common BH Significant Genes Across Objectives - Displaying IFI6, HP, and IFI27 as the top three BH significant genes with logFC > 1, this figure shows the commonality of these genes among ND-0-10, ND-3-10, and NDL-0-strong comparisons.*

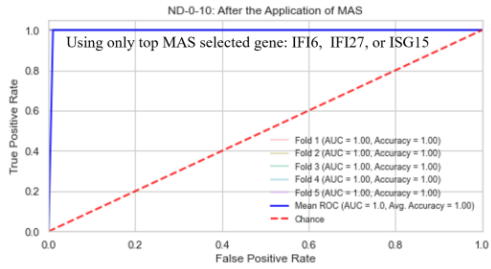
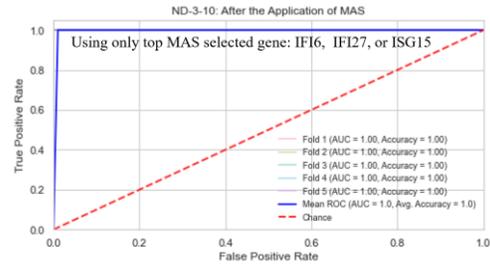

***Figure 5(d).*** *ROC AUC for Classifying NHPs Using SMAS Predictors - This graph illustrates the ROC AUC for classifying NHPs as negative (RT-qPCR on day 0) or positive (day 10) using IFI6, IFI27, or ISG15 (only one gene as the sole predictor in SMAS.*

***Figure 5(e).*** *ROC AUC for Classifying NHPs Using SMAS Predictors - This graph illustrates the ROC AUC for classifying NHPs as negative (RT-qPCR on day 0) or positive (day 10) using IFI6, IFI27, or ISG15 (only one gene) as the sole predictor in SMAS.*

***Figure 5.*** *Gene Expression Analysis and Classification in Objective 1.3*

## Objective 1.4. Comparison of All Negative vs. All Positive Samples (All-N-P):

**Figure 6** presents the results related to objective 1.4. **Figure 6(a)** displays a volcano plot that emphasizes the BH significant genes with a log fold change (logFC) greater than 1, ranked by MAS, as we contrasted all positive NHPs against all negative NHPs for the RT-qPCR test. **Figure 6(b)** illustrates the ROC AUC for 5-fold stratified cross-validation using only the top gene, OAS1, as the predictor for the Simplified MAS (SMAS). **Figure 6(c)** depicts the visualization of all samples using the first three principal components (PC1, PC2, and PC3) as axes. Meanwhile, **Figure 6(d)** shows the 3D visualization of all samples using the top three MAS-selected genes.

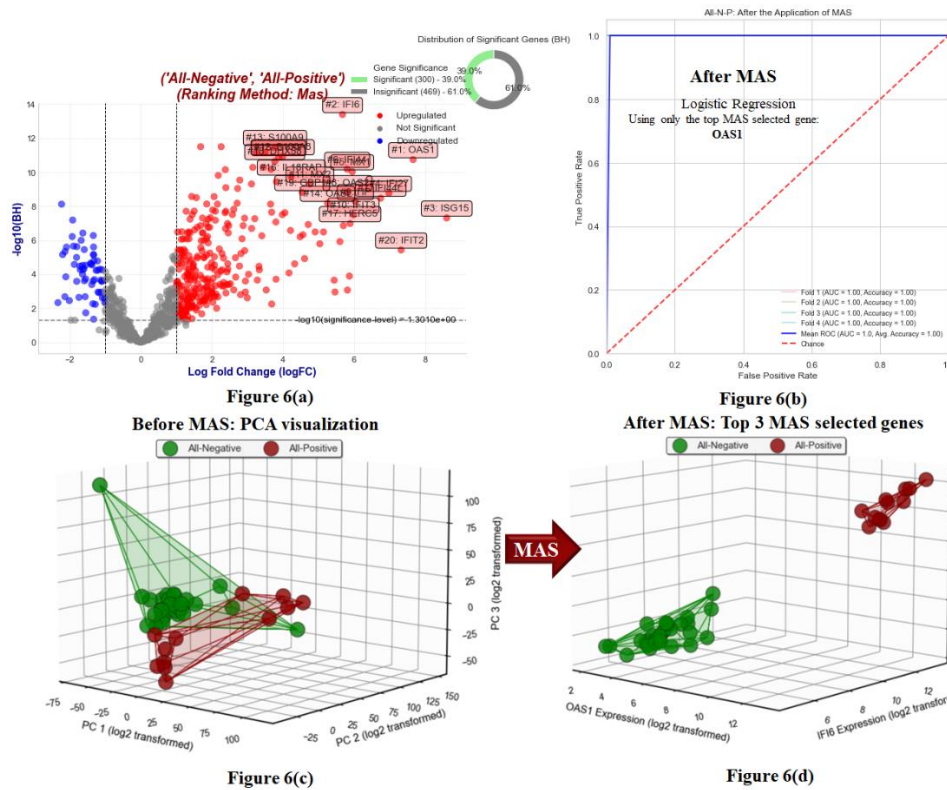

***Figure 6.*** *Analysis and Visualization of Gene Expression for Objective 1.4: Figure 6(a) shows a volcano plot contrasting all positive versus negative NHPs in RT-qPCR, highlighting BH significant genes with logFC > 1 using MAS ranking. Figure 6(b)*

*illustrates the ROC AUC from a 5-fold stratified cross-validation using the top gene, OAS1, in the Simplified MAS (SMAS) approach. Figure 6(c) depicts the visualization of all samples using the first three principal components (PC1, PC2, PC3). Finally, Figure 6(d) presents a 3D visualization of all samples using the top three MAS-selected genes, offering insights into their spatial distribution and expression patterns.*

**Objective 2. Application of MAS Insights for EBOV Classification**

**Objective 2.1. Classification Using Top Selected MAS Genes:**

**Figure 7** demonstrates the results related to Objective 2. Since it was challenging to display all common and unique selected MAS genes for All-N-P, NDL-0-strong, ND-0-10, and ND-3-10, **Figure 7(a)** only shows the common and unique genes for the first three groups. However, the top-3 genes for common genes among all four groups (All-N-P, NDL-0-strong, ND-0-10, and ND-3-10) are IFI6, IFI27, and MX1, respectively. **Figures 7(b)** and **7(c)** illustrate the expression of these top 3 selected genes through a 3D visualization and heat map clustering, respectively. **Figure 7(c)** reveals the hierarchical clustering where the "Ward" method is employed [14]. This agglomerative clustering approach starts with each sample as a separate cluster and iteratively merges them into larger ones. The Ward method aims to minimize the total within-cluster variance, thereby tending to create more evenly sized, spherical clusters. Using these three genes, the method effectively separates the positive and negative groups from each other.

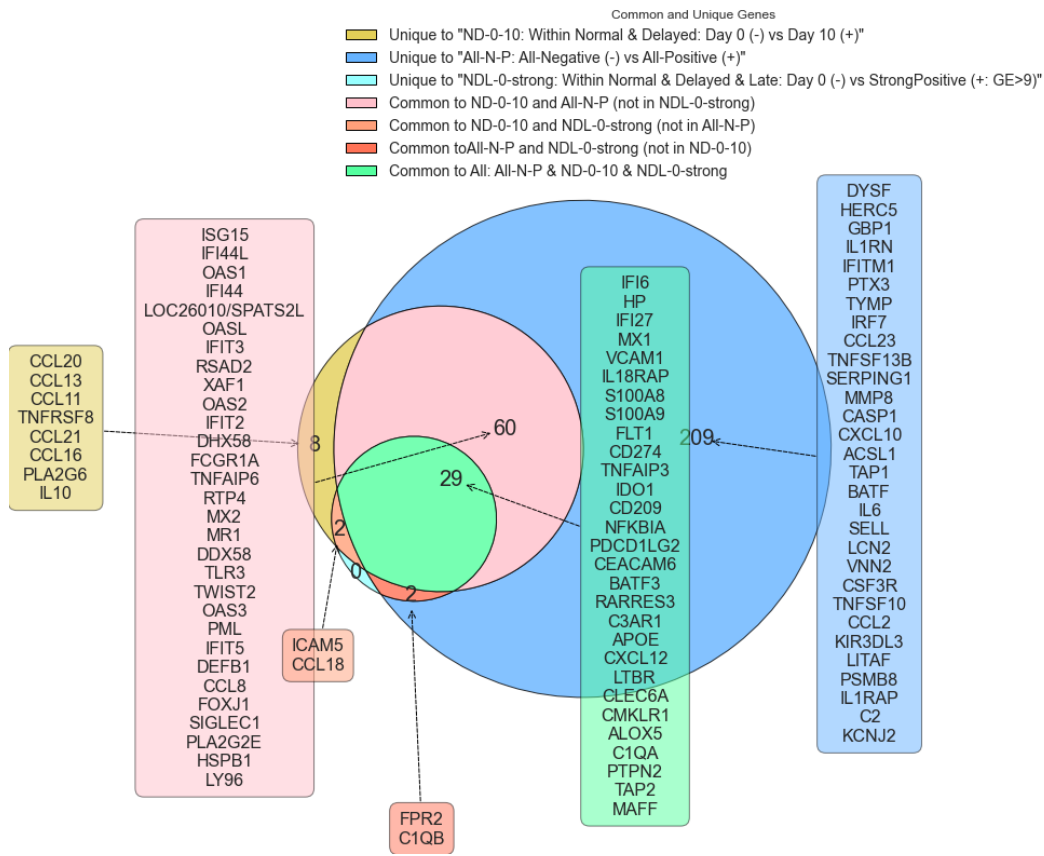

*Figure 7(a).* The figure focuses on the selected MAS genes for All-N-P, NDL-0-strong, ND-0-10, and ND-3-10. Due to the complexity of presenting all common and unique genes across these groups, this figure specifically illustrates only those for the first three groups. However, it highlights IFI6, IFI27, and MX1 as the top three common genes shared among all four groups, underscoring their significance across various analysis categories.

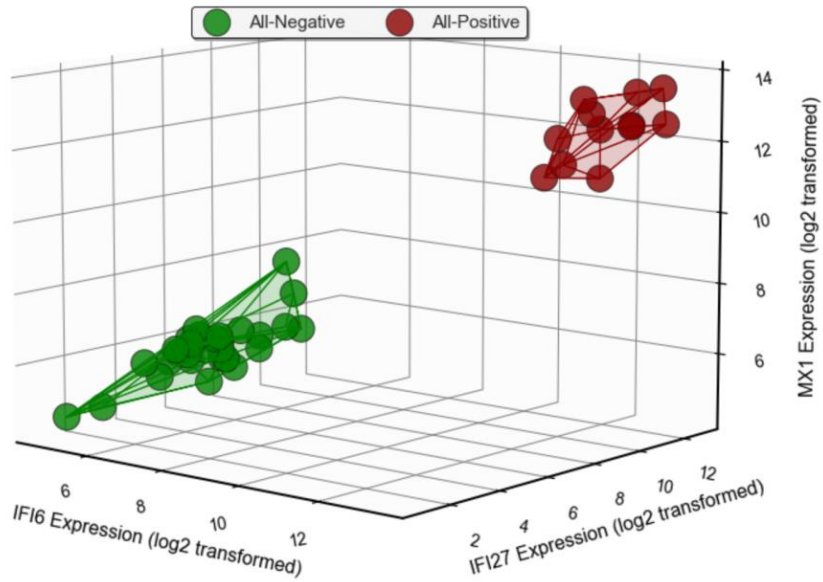

***Figure 7(b)***. *The figure presents a 3D visualization of gene expression, focusing on the top three selected genes: IFI6, IFI27, and MX1.*

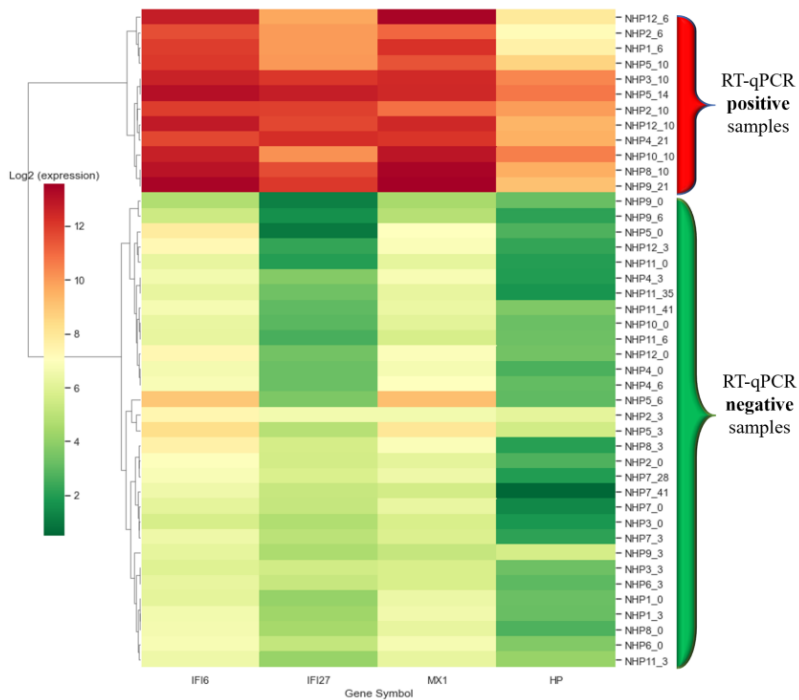

***Figure 7(c)***. *The figure displays a heat map with hierarchical clustering of the top three selected genes: IFI6, IFI27, MX1, and HP.*

***Figure 7***. *Gene Expression Analysis and Predictive Modeling in Objective 2.*

When we repeat this process using the traditional gene ranking commonly used (genes with $\text{logFC} > 1$ and then prioritizing based on BH adjusted p-values, so that the gene with the minimum BH adjusted p-value and $\text{logFC} > 1$ becomes the top gene in the traditional method), it turns out that the top gene is FLT1. Our aim is to pinpoint a gene capable of differentiating between positive and negative RT-qPCR NHP samples.

When focusing on the extreme cases within NDL-0-strong, FLT1 emerges as the top-ranked gene using the traditional ranking method, yet it falls to rank 16 when assessed with the MAS score. While FLT1 proves to be a reliable gene marker within the context of extreme cases in NDL-0-strong, its effectiveness wanes in other scenarios. **Table 1**, which demonstrates the performance of logistic regression via 4-fold cross-validation for All-N-P, indicates that FLT1 is not the most suitable gene for differentiating between negative and positive samples. Using IFI6 as the only predictor and logistic regression as the modeling approach, coupled with 5-fold stratified cross-validation, we attained 100% accuracy and an AUC of 100% for objectives 1.1, 1.3 and 1.4 using balanced data.

*Table 1. Comparative Analysis of Gene Selection Methods for RT-qPCR NHPs. The table highlights the efficacy of the MAS method in identifying IFI6 as a gene capable of perfectly distinguishing between all negative and positive samples, achieving 100% accuracy and AUC. This contrasts with the traditional method, which does not yield such definitive separation. Notably, due to the imbalance in the All-N-P dataset, we randomly selected 12 negative samples to form a balanced dataset comprising 24 samples (12 per class). Following this, we conducted 4-fold stratified cross-validation, repeated 100 times, to calculate the average performance metrics with their standard deviations (STD) for each method.*

| Ranking Method | Top-selected gene | Average AUC | Average Accuracy | Average Precision | Average Recall | Average F1-Score |
|---|---|---|---|---|---|---|
| **MAS (ours)** | **IFI6** | **1.00 ± 0.00** | **1.00 ± 0.00** | **1.00 ± 0.00** | **1.00 ± 0.00** | **1.00 ± 0.00** |
| Traditional (EdgeR [9] or DESeq2 [10]) | FLT1 | 0.88 ± 0.07 | 0.82 ± 0.05 | 0.87 ± 0.06 | 0.80 ± 0.06 | 0.81 ± 0.06 |

**Objective 2.2. Implementing MAS on an Independent Set for Subsequent Supervised Classification:**

To clearly illustrate the differences between traditional ranking methods and the MAS approach, **Figure 8** compares the top 20 genes selected by each method on ND-0-strong. Following the strategy outlined in Section 2 under Objective 2.2, the gene identified as IFI27 emerges as the top selection, $G_{MAS}$, from the MAS process. Utilizing IFI27 as the primary predictor in logistic regression, we conducted 4-fold cross validation with all negative and positive samples, excluding those from ND-0-strong, as shown in **Figure 2** (Objective 2.2). This process was also replicated using the traditional ranking method for comparison. **Table 2** presents the performance of the logistic regression model, comparing the results obtained using the top-selected genes from both MAS and the traditional method.

*Figure 8. Comparative Analysis of Gene Selection between Traditional Ranking and MAS Methods for Objective 1.1. The figure displays the top 20 genes selected by the traditional ranking method (left) and the MAS method (right) for objective 1.1. It highlights differences in gene prioritization between the two methods, revealing potential insights into gene selection for distinguishing negative and positive samples in RT-qPCR NHPs.*

*Table 2. Performance comparison of logistic regression model using top genes selected by MAS and the traditional methods from ND-0-strong, applied to remaining samples excluding ND-0-strong.*

| Ranking Method | Top-selected gene | Average AUC | Average Accuracy | Average Precision | Average Recall | Average F1-Score |
|---|---|---|---|---|---|---|
| **MAS (ours)** | **IFI27** | **1.00 ± 0.00** | **1.00 ± 0.00** | **1.00 ± 0.00** | **1.00 ± 0.00** | **1.00 ± 0.00** |
| Traditional (EdgeR [9] or DESeq2 [10]) | FLT1 | 0.72 ± 0.27 | 0.69 ± 0.17 | 0.71 ± 0.22 | 0.72 ± 0.14 | 0.67 ± 0.18 |

# 4. Discussion

In our study, the comprehensive differential expression analysis using MAS elucidated significant changes in gene expression due to EBOV infection in nonhuman primates (NHPs). Below, we discuss the results obtained in Section 3 for different objectives.

**Objective 1. Comprehensive Differential Expression Analysis Using MAS Scoring:**

**Objective 1.1. Strong Negative vs. Strong Positive Analysis (NDL-0-strong):**

The "Strong Negative vs. Strong Positive Analysis (NDL-0-strong)" within our study has proven to be a critical component in discerning the genetic alterations associated with extreme cases of EBOV infection. Utilizing the MAS method, we successfully identified key genes, notably IFI27, IFI6, and HP, which displayed significant expression level changes in strong positive cases as opposed to strong negatives. This analysis, enriched by three-dimensional visualization as illustrated in **Figure 3(b)** and enhanced by logistic regression with 5-fold stratified cross-validation (see **Figure 3(c)**), underscores the strength of our approach.

The precision and effectiveness of our model are most notably reflected in the ROC curves, which exhibit a remarkable 100% AUC and accuracy metrics. This exceptional performance is largely attributed to our focused reliance on a single, highly indicative gene, IFI27. Such a focused approach not only addresses the complexity of EBOV pathogenesis but also reveals a clear genetic signature. This signature is of paramount importance, presenting itself as an instrumental asset for the future development of diagnostic and therapeutic strategies targeting EBOV infections. The attainment of 100% accuracy and AUC in our analysis highlights the potential of our methodological approach in revolutionizing the understanding and management of severe viral infections.

Supporting this selection of IFI27, several studies have provided insights into its role in different viral infections. Huang et al. [15] highlighted IFI27 as a gene associated with the progression of HIV infection, suggesting its potential as a target for immunotherapy. Similarly, Shojaei et al. [16] emphasized the expression of IFI27 in the respiratory tract of COVID-19 patients, correlating its increased expression with a high viral load. Their findings also underscored the role of IFI27 as a marker of systemic host response, demonstrating its high sensitivity and specificity in predicting clinical outcomes. In the context of influenza, Tang et al. [17] identified IFI27 as a single-gene biomarker with high predictive accuracy for distinguishing between influenza and bacterial infections. Their work showed that IFI27 is upregulated by TLR7 in plasmacytoid dendritic cells, more responsive to influenza virus than bacteria, and confirmed its expression in influenza patients through multiple patient cohorts.

Further corroborating the relevance of IFI27, Villamayor et al. [18] explored its role in regulating innate immune responses to viral infections. Their study highlighted a novel function of IFI27 in modulating responses triggered by cytoplasmic RNA recognition and binding. The interaction of IFI27 with nucleic acids and the PRR retinoic acid-inducible gene I (RIG-I) was a key discovery, revealing its potential to impair RIG-I activation and thus modulate innate immune responses.

The convergence of findings from various studies establishes IFI27 as a critical gene in the context of viral infections, highlighting its significant role across different viral diseases. These insights further validate the selection of IFI27 through the MAS method, particularly within the NDL-0-strong objective for EBOV infection in NHPs. The prominence of IFI27 in this analysis underscores its importance in deciphering the complex mechanisms of viral pathogenesis and the associated immune responses. This validation by MAS in identifying IFI27 is significant, as it indicates the gene's involvement in crucial biological pathways during severe viral infections. By focusing on IFI27, our study not only identifies a key genetic marker but also paves the way for a deeper understanding of the interaction networks and pathways it influences or participates in. This comprehensive approach enriches our grasp of the genetic underpinnings at play in viral infections and could potentially inform the development of targeted therapeutic interventions. The pathway analysis of IFI27, therefore, offers a more nuanced view of its role, extending beyond mere gene expression to its functional implications in the context of viral infections and immune system responses.

**Objective 1.2. Differential Expression in Group 1 (N-0-6 & N-3-6):**

As the positive RT-qPCR tested NHPs on day 6 post-infection are contrasted against the same NHPs on days 0 and 3 post-infection, when they were negative for RT-qPCR in Group 1, IFI27 emerges as the top MAS-selected gene (see **Figure 4**). Although we did not perform supervised prediction due to limited samples, **Figure 4(e)** suggests that IFI27 perfectly separates the positive and negative samples. Considering IFI27 as the top MAS-selected gene for both Objectives 1.1 (extreme cases) and 1.2 (regular cases), it becomes evident that IFI27 is a signature marker for Ebola infection.

The findings of our study, particularly the identification of IFI27 as a key biomarker in the early stages of EBOV infection, align closely with the insights from Normandin et al. [19]'s comprehensive analysis. Their work, which involved extensive RNA sequencing across multiple tissues in rhesus monkeys infected with EBOV (EBOV), revealed a significant correlation between the expression of IFI27 and viral RNA load. This parallel discovery in both studies underscores the crucial role of IFI27 in the host's immune response to EBOV infection. By demonstrating the early detection capability of IFI27 as a biomarker in our study and its correlation with viral load in various tissues in the study by Normandin et al., a more nuanced understanding of the molecular mechanisms of EVD pathogenesis emerges.

Blengio et al. [20] conducted a study exploring the gene expression profiles in response to Ebola vaccination, revealing crucial insights into the immune response mechanisms. Among the key findings, IFI27 stood out, being one of the early expressed genes whose upregulation on day 1 post-vaccination correlated with the magnitude of the antibody response observed both 21 days after the MVA-BN-Filo and 364 days after the Ad26.ZEBOV vaccinations.

Kash et al. [21] conducted an in-depth analysis of the immune response in a patient with severe EVD, revealing critical insights into the dynamics of host gene expression in relation to the course and severity of the illness. Their study, conducted during the unprecedented 2013–2015 Ebola outbreak in Guinea, Liberia, and Sierra Leone, involved daily microarray analysis of peripheral blood samples from a patient treated at the National Institutes of Health Clinical Center. This comprehensive approach allowed them to correlate gene expression changes with various clinical parameters, including viral load, antibody responses, coagulopathy, organ dysfunction, and eventual recovery.

A key finding from Kash et al. [21]'s research was the identification of IFI27 as a significantly expressed gene in response to EBOV replication. Their analysis revealed that IFI27, along with other type I interferon-stimulated genes (ISGs), showed marked expression changes correlating with phases of the disease. Particularly on day 13 post-infection, IFI27, among other ISGs, was found to be highly expressed, aligning with the peak antiviral response in the patient. This aligns with our findings and those of other studies like Normandin et al. [19] and Blengio et al. [20], where IFI27 was identified as a critical biomarker in the context of EBOV infection and response.

As discussed in Objective 1.1, IFI27's role extends beyond being a mere biomarker. The consistent upregulation of IFI27 in Ebola-infected NHPs, as compared to their negative counterparts, underlines its significance in the pathophysiology of the infection. This pattern indicates a potential role of IFI27 in the host's response to the EBOV, possibly in the modulation of the immune system or the activation of specific cellular pathways. Given the crucial role of interferon-stimulated genes like IFI27 in the innate immune response, its significant expression in Ebola-infected NHPs could reflect an activation of defense mechanisms against the virus. This aligns with findings in other viral infections, where IFI27 is implicated in immune response modulation and viral replication control.

Therefore, the prominence of IFI27 in Ebola infection, at least in NHP models, highlights its potential as a therapeutic target or a biomarker for early diagnosis and treatment strategies. Understanding the exact mechanisms by which IFI27 influences Ebola pathogenesis could pave the way for novel approaches to manage and treat this severe viral infection. This study opens new avenues for investigating IFI27's role in EBOV infection and its potential application in clinical settings.

**Objective 1.3. Analysis Within Normal and Delayed Groups (ND-0-10 & ND-3-10):**

When we contrasted the RT-qPCR positive NHPs on day 10 against the same NHPs on days 0 and 3, when they were still negative in the RT-qPCR test, interesting results emerged. Figure 5(a) indicates that from day 0 to day 10, a total of 99 genes are significantly upregulated (as indicated by Benjamini-Hochberg significance) with a log fold change (logFC) greater than 1. Furthermore, from day 3 to day 10, only 37 of these genes remained significantly upregulated. The difference in gene expression between day 3 and day 10 is particularly noteworthy. While the number of significantly upregulated genes decreases to 37, these genes (e.g., ISG15, IFI6, IFI44L, HP, OAS1, IFI44, OASL, and IFI27) likely play a crucial role in the response to the progressing infection. The reduction in the number of upregulated genes could indicate a more targeted or evolved response of the host's immune system as the infection progresses. Alternatively, it may reflect a shift in the virus-host interaction dynamics over time. This pattern of gene expression highlights the complex and evolving nature of the host response to EBOV infection. Identifying these key genes that remain upregulated over time could provide insights into critical pathways and mechanisms that the virus exploits or the host deploys in response to the infection. This understanding could be invaluable for developing targeted therapies or diagnostics for the EBOV.

**Figure 5(b)** demonstrates the common and unique MAS-selected genes with logFC greater than 1 among the negative RT-qPCR results on day 0 compared to the positive ones on day 10 and day 6 and the strong positives on days 10 or 21. It turns out that the only Benjamini-Hochberg (BH) significant gene with logFC greater than 1, common among all these three groups, is IFI27. When the N-0-6 group is replaced with ND-3-10, there are 12 BH significant genes with logFC greater than 1, with IFI6, HP, IFI27, and MX1 as the top selected genes (see **Figure 5(c)**). Using either IFI6, IFI27, or ISG15 as the single predictor for logistic regression, we achieved 100% accuracy in separating the positive NHPs on day 10 from their counterparts on days 0 or 3 (see **Figure 5(d-e)**). This analysis provides crucial insights into the gene expression profiles in NHPs associated with EBOV infection. IFI27's consistent upregulation across different groups and time points indicates its potential as a biomarker for Ebola infection. Its presence in all examined groups underscores its importance in early virus detection. The ability to differentiate between positive and negative samples with 100% accuracy using IFI6, IFI27, or ISG15 as predictors highlights the significant alteration of these genes following infection and their potential as reliable indicators of EBOV presence in NHPs.

Building on this, several studies have explored the roles of related genes in viral infections. Sajid et al. [22] focused on IFI6 in HBV, demonstrating its antiviral properties post-type I IFN-alpha stimulation. Similarly, Qi et al. [23] showed that IFI6 expression increases in DENV-infected cells, influencing apoptosis-related processes. Liu et al. [24] studied lncRNA-IFI6 in HCV infection, uncovering its regulatory effects on the antiviral ISG IFI6. Park et al. [25] investigated IFI6's genetic variations in chronic liver disease patients with HBV.

Morales et al. [26] reviewed the current understanding of ISG15, examining its role in mediating protection against different viral infections and exploring the mechanisms by which it exerts antiviral activity. Their work highlights the importance of ISG15 in the immune system's response to viral threats, emphasizing its potential as a target for therapeutic interventions in viral diseases. Jeon et al. [27] focused on the role of ISG15 in the immune response, particularly its involvement in the conjugation process known as ISGylation. Their findings indicate ISG15's significant role in the innate immune response. Perng et al. [28] focused their review on the multifaceted role of ISG15, a ubiquitin-like protein, in the host's response to viral infection. They discussed how ISG15, induced by type I interferons, not only directly inhibits viral replication but also modulates various host responses, including damage repair, immune response, and other signaling pathways. Their review highlighted the diverse and pathogen-dependent actions of ISG15, emphasizing its importance in antiviral defense. Additionally, they explored how viruses evolve strategies to evade ISG15's actions. The review also integrated new findings on individuals deficient in ISG15 and the identification of a cellular receptor for ISG15, providing deeper insights into how ISG15 influences the host response to viral infections.

These studies collectively enhance our understanding of the roles of genes like IFI6 and ISG15 in the body's response to viral infections, providing valuable context for our findings on EBOV. The mechanisms elucidated in these studies, particularly relating to IFI6's antiviral properties and ISG15's role in immune modulation and viral replication inhibition, offer insights that may be relevant in understanding the specific genetic responses we observed in NHPs infected with Ebola.

**Objective 1.4. Comparison of All Negative vs. All Positive Samples (All-N-P):**

When contrasting all RT-qPCR positive tested NHPs against all negative ones, regardless of time, OAS1 emerged as the top MAS-selected gene, perfectly separating negative from positive samples with 100% accuracy (see **Figure 6**). The second and third top MAS-selected genes are IFI6 and ISG15, consistent with our findings in previous objectives.

Melchjorsen et al. [29] investigated the expression patterns of the 2′-5′ oligoadenylate synthetase (OAS) family genes, particularly Oligoadenylate synthetase 1 (OAS1) and OASL, during viral infections such as with Sendai virus and Influenza A virus. Their findings suggest that OASL behaves like an antiviral gene, providing new insights into the roles of the OAS gene family members in the immune response to viral infections. Fish et al. [30] conducted a comprehensive study on OAS1, highlighting its critical role as a first line of defense against various viral pathogens, particularly in Old World monkeys. Their research revealed that OAS1 is evolving under positive selection in these species, with most of the positively selected sites located in the RNA-binding domain (RBD), which is responsible for binding viral dsRNA. This positive selection, especially concentrated in a specific region of the RBD, suggests a sub-functionalization within this domain, potentially enhancing OAS1's ability to recognize and bind diverse viral RNAs. Additionally, Fish et al. [30] identified positively selected residues around the active site's entry, indicating an evolutionary adaptation to possibly evade viral antagonism or to produce oligoadenylates of varying lengths. These findings underscore the evolutionary pressures shaping OAS1 and its importance in the innate immune response to viral infections.

Building upon these studies, OAS1's pathway in viral infection response can be delineated. OAS1 is activated upon recognizing double-stranded RNA (dsRNA) from viruses, including Ebola, leading to the synthesis of 2′-5′-linked oligoadenylates (2-5A). This synthesis subsequently activates Ribonuclease L (RNase L), which then cleaves both viral and host RNA, effectively inhibiting viral replication. However, this activation by OAS1 is a double-edged sword, as indicated by Carey et al. [31]'s findings show that OAS1 activation can also lead to autoactivation by host RNAs, presenting potential costs to the host. The evolutionary study by Fish et al. [30] added another layer to this pathway by suggesting that OAS1's RNA-binding domain has evolved under positive selection in some primate species, enhancing its ability to recognize and bind diverse viral RNAs. This evolutionary adaptation may be a response to the need for balancing effective viral defense with the minimization of collateral damage to the host. In the context of EBOV infection, this pathway underscores OAS1's critical role in the initial detection and response to the virus, marking it as a key player in the host's antiviral defense mechanism.

**Objective 2. Application of MAS Insights for EBOV Classification**

**Objective 2.1. Classification Using Top Selected MAS Genes:**

Objective 2.1 of our study focused on identifying a definitive gene marker for distinguishing between positive and negative NHPs in the RT-qPCR test, irrespective of the severity or level of Genome Equivalent. Our approach involved selecting the top MAS-selected gene with a logFC greater than 1, consistently identified across objectives 1.1, 1.2, and 1.4. It turns out that IFI6, IFI27, and MX1 are the top common genes across all objectives (**Figure 7(a)** shows the common genes across three groups). According to **Figures 7(b)** and **7(c)**, using any of these genes effectively separates the positive and negative samples into distinct clusters.

To evaluate the effectiveness of the MAS ranking method, we compared it with the traditional ranking method. We found that the top MAS-selected gene, IFI6, successfully differentiated all negative from positive samples with 100% average accuracy and AUC. In contrast, the top gene selected by the traditional ranking method, FLT1, achieved an average accuracy of 82% and an average AUC of 0.88, as indicated in **Table 1**. Additionally, we compared the MAS and traditional ranking mechanisms within the extreme cases of Objective 1.1. As shown in **Figure 8**, the MAS approach effectively accounts for both biological and statistical significance, highlighting its superiority in identifying key genetic markers in EBOV infection studies.

We have discussed the pathways of IFI6 and IFI27 in previous objectives. Regarding MX1, Caballero et al. [32] focused on the host response to EBOV infection, revealing that MX1, along with ISG15 and OAS1, is significantly upregulated in circulating immune cells during infection. Their study found that these genes, particularly MX1, are part of a strong innate immune response triggered by active virus replication. This response, which also includes other interferon-stimulated genes, contrasts with in vitro evidence suggesting suppression of innate immune signaling. The prominence of MX1 in the immune response to the EBOV highlights its potential role in the pathogenesis of the infection and underscores its significance as a key component of the host's defense mechanism against viral threats. Pillai et al. [33] highlighted the critical role of MX1 in the immune response to viral infections (particularly to IAV) and underscored the complex interplay between viral defense mechanisms and subsequent bacterial complications. Fuchs et al. [34] investigated the role of MX1 proteins in bats, particularly in relation to their innate immune defense against various viruses, including Ebola. Bats, known reservoirs for zoonotic viruses such as Ebola, typically do not exhibit clinical symptoms from these infections. They focused on cloning MX1 cDNAs from three bat families and analyzing their antiviral potential. They found that bat MX1 proteins are key factors in controlling viral replication, including Ebola, in their bat hosts, and offer insights into the coevolution of these proteins with bat-borne viruses.

**Objective 2.2. Implementing MAS on an Independent Set for Subsequent Supervised Classification:**

The results presented in **Table 2** underscore the remarkable efficiency of the MAS method in gene selection for predictive modeling, especially when contrasted with traditional ranking methods. The perfect scores across all metrics achieved by the logistic regression model using the MAS-selected gene IFI27, including Average AUC, Accuracy, Precision, Recall, and F1-Score, firmly establish MAS as a highly effective approach in this context. These scores not only reflect the method's impeccable accuracy in classifying RT-qPCR positive and negative samples but also its consistency, as indicated by the absence of variability in the results. This level of performance highlights the potential of MAS to reliably identify genes that are not just statistically significant but also practically relevant for diagnostic and therapeutic applications, particularly in the field of EBOV research.

In stark contrast, the traditional ranking method, as evidenced by its moderate scores using the gene FLT1, demonstrates a less robust performance. While it maintains a certain level of effectiveness, the considerable variability in its metrics suggests lower reliability and precision. This comparison draws attention to the limitations inherent in traditional methods, which often prioritize statistical significance over practical utility. The discrepancies in performance between the two methods may stem from their differing focuses: MAS balances statistical and biological significance, while traditional methods may disproportionately emphasize statistical metrics, potentially overlooking genes with substantial biological relevance.

In summary, a key finding from our analysis was the identification of IFI27 as a prominent gene in both extreme and regular cases of EBOV infection. IFI27 consistently emerged as the top MAS-selected gene, establishing itself as a robust marker for the presence of the virus. This was further corroborated by its exceptional performance in logistic regression models, where it demonstrated high accuracy in differentiating between positive and negative samples. Moreover, our study revealed the significant role of genes like IFI6, which, alongside IFI27, showed marked expression level changes in response to the infection. These genes formed a distinct pattern of expression, aligning with the progression and severity of the infection in NHPs.

We also observed that other genes, such as MX1 and ISG15, were significantly upregulated, suggesting their involvement in the host's immune response to the EBOV. The pattern of upregulation in these genes provided insights into the complex interplay between the host's defense mechanisms and viral pathogenesis. In a broader classification analysis, we found that genes like OAS1 also played a significant role in our study. OAS1 was identified as a crucial marker, effectively separating positive from negative EBOV samples with high accuracy. This underscores its potential role in the innate immune response and its importance as part of the genetic signature of EBOV infection.

Our findings highlight the power of the MAS method in unraveling the complex genetic responses to EBOV infection. By identifying key genes that are significantly altered during infection, our study not only contributes to the understanding of EBOV pathogenesis but also paves the way for developing targeted diagnostic and therapeutic strategies. The distinct gene signatures we identified could serve as crucial tools in managing and treating EBOV infections, providing a foundation for future research in this critical area of virology.

## 5. Conclusions and Study Limitations

This research represents a significant leap forward in understanding and detecting EBOV infection in nonhuman primates. By implementing the novel Supervised Magnitude-Altitude Scoring (SMAS) methodology and leveraging NanoString gene expression data, we have successfully pinpointed pivotal biomarkers. Notably, genes like IFI6 and IFI27 emerged as crucial, demonstrating remarkable accuracy and AUC metrics in categorizing various stages of EBOV infection. The study not only pinpointed specific genes that undergo significant expression changes in response to the EBOV but also illuminated the broader impact of these changes on the immune response. The increased expression of genes such as MX1 and ISG15 underscores the intricate dynamics of host defense mechanisms in combating viral pathogenesis. These insights are instrumental for the virology field, laying the groundwork for more refined diagnostic methods and therapeutic approaches against EBOV infection.

However, the study's scope was limited to a small group of nonhuman primates. Expanding the sample size and diversity could yield more robust insights and further validate our findings. While promising, the applicability of these results to human populations remains uncertain, necessitating further investigation to verify the effectiveness of these biomarkers in humans. Additionally, our study did not consider the potential influence of environmental or external factors on gene expression and disease progression in nonhuman primates. Future studies should also explore longitudinal and functional aspects to fully comprehend the roles of these genes in Ebola pathogenesis and the broader implications of our findings. Addressing these limitations will enhance our understanding and contribute to more effective strategies for managing EBOV infections.

## 6. Data Availability

The NanoString gene expression data utilized in this study are derived from the work of Speranza et al. [1], which can be accessed in accordance with the permissions and guidelines provided by the original authors.